\documentclass[preprint]{revtex4}
\usepackage{graphicx}
\usepackage{amssymb}
\newcommand{\be}{\begin{eqnarray}}
\newcommand{\ee}{\end{eqnarray}}

\begin{document}

\title{Generalized Parton Distributions for the Proton in Position Space :
Zero Skewness}
\author{\bf D. Chakrabarti$^a$, R. Manohar$^b$, A. Mukherjee$^b$}
\affiliation{$^a$ Department of Physics, 
Indian Institute of Technology Kanpur,
Kanpur 208016, India\\
$^b$ Department of Physics,
Indian Institute of Technology Bombay,\\ Powai, Mumbai 400076,
India.}
\date{\today}

\begin{abstract}
We present a study of the parton distributions in transverse position or 
impact parameter space using a recent parametrization of u and d quark 
generalized parton distributions (GPDs) $H(x,t)$ and $E(x,t)$ at zero
skewness. We make a comparative study between different
parametrizations and discuss the region of validity of the positivity
condition.   
\end{abstract}
\maketitle


\vskip0.2in
\noindent
{\bf Introduction}

Generalized parton distributions (GPDs) contain a wealth of information
about the nucleon structure (see \cite{rev} for example). Unlike the
ordinary parton distributions (pdfs) which at a given scale depend only 
on the longitudinal momentum fraction $x$ of the parton, GPDs are functions
of three variables, $x$, $\zeta$ and $t$ where the so-called skewness 
$\zeta$ gives the
longitudinal momentum transfer and $-t$ is the square of the momentum
transfer in the process. These are called off-forward parton distributions. 
The GPDs give interesting information about the spin and orbital angular
momentum of the quarks and gluons in the nucleon as well as their spatial
distribution. They are experimentally accessed through the overlap of
deeply virtual Compton scattering (DVCS) and Bethe-Heitler (BH) process as
well as exclusive vector meson production \cite{rev}. Data has been obtained
at HERA collider, by the H1 \cite{H1,H2} and ZEUS \cite{ZEUS1,ZEUS2} collaborations 
and  HERMES \cite{HERMES} fixed target
experiment. DVCS experiments are also
being done at JLAB Hall A and B \cite{CLAS}. COMPASS at CERN has programs 
to access GPDs through muon beams \cite{COMPASS}. Experimental observables, 
however involve a convolution of GPDs, and so modeling GPDs is interesting. GPDs 
reduce to ordinary forward parton
distributions (pdfs) in the forward limit. Moments over $x$ give nucleon
form factors. These act as useful constraints to model the GPDs.
A large number of models or parametrizations
have been proposed for GPDs. Here we do not plan to  review all of them
but mention only those that are relevant for us. A detailed list of the main
lines of approach and their present status with respect to the data
 can be found in \cite{boffi}. Moments of GPDs have been calculated 
on lattice as well. In \cite{marc}, GPDs at
zero skewness $\zeta$ have been parametrized in a Regge type model at small
$x$ and using a modified Regge  parametrization at large $x$ and $t$.           
It was found to describe the basic features of proton and neutron
electromagnetic form factors. Similar small $x$ Regge type behaviour was
used in the modeling of pion GPDs in \cite{rad}.
Another parametrization of GPDs at zero skewness is given in \cite{diehl}.
Here a Regge motivated $x$ dependence at small $x$ was interpolated to large
$x$ region. The GPDs at the input scale were fitted to the experimental data
on Dirac and Pauli form factors. An exponential $t$ dependence was used. The
form of the $t$ dependence was found to be unchanged by the scale evolution.
In \cite{liuti} a recent parametrization was proposed for zero skewness
and an extension to non-zero $\zeta$ was done in \cite{l2}. At the input
scale the GPDs were parametrized by a spectator model term multiplied by a
Regge motivated term. The parameters were obtained by fitting the forward
pdfs and form factors. 

For non-zero $\zeta$, the GPDs have to satisfy an additional constraint,
namely polynomiality. In certain models, for example using the overlap of
light-front wave functions (LFWFs), it is very difficult to obtain a suitable
parametrization of the higher Fock components of the wave function in order
to get the polynomiality of GPDs. Polynomiality is
satisfied by construction only if one considers the LFWFs of simple spin 
$1/2$ objects like a dressed quark or a dressed electron  in perturbation
theory instead of the proton \cite{overlap,us}. A recent fit to the DVCS 
data at
small Bjorken $x$ from H1 and ZEUS was done in \cite{dieter}, using the
conformal Mellin-Barnes representation of the DVCS amplitude. However, to
get the GPDs one has to do an inverse Mellin transform, and a knowledge of
all moments are required for that.  

At zero skewness $\zeta$, if one performs a Fourier transform (FT) of the 
GPDs with respect to the momentum transfer in the transverse direction 
$\Delta_\perp$, one gets the so called impact parameter dependent parton 
distributions
(ipdpdfs), which gives how the partons of a given longitudinal momentum are
distributed in transverse position (or impact parameter $b_\perp$) space.
These obey certain positivity constraints and unlike the GPDs themselves,
have probabilistic interpretation \cite{burkardt}. These give an 
interesting interpretation of Ji's angular momentum sum rule \cite {ji}. Due
to rotational invariance, the same relation should hold for all components
of the angular momentum $\vec{J_q}$. In the impact parameter space,  
the relation for $J^\perp_q$ has a simple partonic interpretation for
transversely polarized state 
\cite{bur05};  the term containing $E(x,0,0)$ arises due to a transverse 
deformation of the GPDs 
in the center of momentum frame. The term containing $H(x,0,0)$ is an
overall transverse shift when going from the transversely polarized state 
in instant form (rest frame) to the front form (infinite momentum frame).
On the other
hand, in \cite{hadron_optics}, real and imaginary parts of the DVCS amplitudes are
expressed in longitudinal position space by introducing a longitudinal
impact parameter $\sigma$ conjugate to the skewness $\zeta$, and it was
shown that the DVCS amplitude show certain diffraction pattern in the 
longitudinal position space. Since Lorentz boosts are
kinematical in the front form, the correlation determined in the
three-dimensional $b_\perp, \sigma$ space is frame-independent.
As GPDs depend on a sharp $x$, the Heisenberg uncertainty
relation restricts the longitudinal position space
interpretation of GPDs themselves. It has, however, been shown
in \cite{wigner} that one can define a quantum mechanical Wigner
distribution for the relativistic quarks and gluons inside the proton.
Integrating over $k^-$ and $k^\perp$, one
obtains a four dimensional quantum distribution which is a function
of ${\vec{r}}$ and $k^+$ where  ${\vec{r}}$ is the quark phase space
position defined in the rest frame of the proton. These
distributions are related to the FT of GPDs in the same frame. This gives a
3D position space picture of the GPDs and of the proton.

In this series of work, we plan to investigate the GPDs for the proton 
in transverse and
longitudinal position space. In this first paper, we study the recently
parametrized form in \cite{liuti} in impact parameter space and make a  
comparative study with other models.


\vskip0.2in
\noindent
{\bf Parametrization of the GPDs} 

We consider the parametrization in \cite{liuti} for the GPDs :

\noindent {\bf Set I} 
\begin{eqnarray}
H^I(x,t) & = & G_{M_{x}^I}^{\lambda^I}(x,t) \,  
 x^{-\alpha^I -\beta_1^I (1-x)^{p_1^I} t}
\label{param1_H}
\\
E^I(x,t) & = & \kappa \, G_{M_x^I}^{\lambda^I}(x,t) \,
x^{-\alpha^I -\beta_2^I (1-x)^{p_2^I} t}
\label{param1_E}
\end{eqnarray}

\noindent {\bf Set II} 
\begin{eqnarray}
H^{II}(x,t)& = & G_{M_x^{II}}^{\lambda^{II}}(x,t) \,
x^{-\alpha^{II} - \beta_1^{II}  (1-x)^{p_1^{II}} t}
\label{param2_H}
\\
E^{II}(x,t) & = & G_{\widetilde{M}_x^{II}}^{\widetilde{\lambda}^{II}}(x,t) \,
x^{-\widetilde{\alpha}^{II} - \beta_2^{II}  (1-x)^{p_2^{II}} t}
\label{param2_E}
\end{eqnarray}

All parameters except for $p_1$ and $p_2$ are flavor dependent.
The function $G$ has the same form for both parametrizations, I and II:

\begin{eqnarray}
G_{M_x}^{\lambda} (x,t) = && 
{\cal N} \frac{x}{1-x} \int d^2{\bf k}_\perp \frac{\phi(k^2,\lambda)}{D(x,{\bf k}_\perp)}
\frac{\phi({k^{\prime \, 2},\lambda)}}{D(x,{ \bf k}_\perp +(1-x){\bf 
\Delta}_\perp)},    \label{gkaava}
\end{eqnarray}
where

\begin{equation}
D(x,{\bf k}_\perp) \equiv  k^2 - m^2,
\end{equation}

\begin{eqnarray}
k^2 & = & x M^2 - \frac{x}{1-x} M_x^{2}  - \frac{{\bf k}_\perp^2}{1-x} \\
k^{\prime \, 2} & = & xM^2 - \frac{x}{1-x} M_x^{2} - \frac{({\bf k}_\perp - (1-x)\Delta)^2}{1-x},
\end{eqnarray}
and 
\begin{eqnarray}
\phi(k^2,\lambda) = \frac{k^2-m^2}{\vert k^2-\lambda^2\vert^2 },
\label{phi}
\end{eqnarray}

Here $\zeta$, the skewness variable is taken to be zero, in other words,
momentum transfer is in the transverse direction. $t$ is the invariant
momentum transfer squared, $t=-\Delta^2$, and $x$ is the fraction of the light
cone momentum carried by the active quark, $k$ being its momentum. 
The mass parameters are  $m$, the struck quark mass, and $M$, the proton mass.
The normalization factor includes the nucleon-quark-diquark coupling, and it
is  set to ${\cal N} = 1$ GeV$^6$. 

The $u$ and $d$ quark contributions to the anomalous magnetic moments are:
\begin{equation}
\kappa \equiv \kappa^q= \left\{ \begin{array}{ll}
\kappa^d = -2.03,& \; {\rm for} \; q=d
\\
\kappa^u/2 = 1.67/2, & \; {\rm for}\; q=u
\end{array} \right. .
\label{kappaq}
\end{equation}

The parameters are listed in \cite{liuti} for both the sets.
The parameters  $M_{x}^q$, $\lambda^q$ and $\alpha^q$, $q=u,d$, obtained
at an initial scale $Q_0^2$ ($Q^2_0 = 0.094$ GeV$^2$),
and they are the same for both Sets I and II, in Set I they are by 
definition the same for the functions $H$ and  $E$ 
(see Eqs.~(\ref{param1_H},\ref{param1_E})).
The parameters in this model are obtained by fitting the experimental data
on the nucleon electric and magnetic form factors (see \cite{liuti} for
reference). For the forward limit, Alekhin \cite{ale} leading order (LO) pdf
sets were fitted within the range $10^{-5} \le x \le 0.8$ and $4 \le Q^2 \le
240 ~  \mathrm{GeV}^2$ by valence distribution and using the baryon number
and momentum sum rules. The input scale $Q_0^2=0.094 ~ \mathrm{GeV}^2$ is
obtained as a parameter in this model. The low value of $Q_0^2$ results from
the requirement that only valence quarks contribute in the momentum sum
rule.

The above phenomenologically motivated parametrization of the GPDs 
$H(x,t)$ and $E(x,t)$ at zero skewness $\zeta$ was done using a
spectator model calculation at the  low input scale. The
spectator model has been used for its simplicity and for the fact that it
is flexible enough to predict the main features of a number of distribution
and fragmentation functions in the intermediate and large $x$ region. The
spectator mass is chosen to be different for different quark flavor GPDs.
However, similar to the case of pdfs, the spectator model is not able to
reproduce quantitatively the small $x$ behaviour of the GPDs. So a
`Regge-type' term has been considered multiplying the spectator model
function $G_{M_x}^\lambda$. The parameters were obtained by fitting the form
factors and forward pdfs. Two versions of the parametrizations were
used and are given by set I and set II. The GPD $E$ is unconstrained by 
the data on forward pdfs, so in set II an additional normalization
condition has been imposed 
\be
\int_0^1 dx E_q(x,t=0)=\kappa^q
\ee
with the experimental values of $\kappa^u$ and $\kappa^d$.

Although $H^u$ and $H^d$ are similar in behaviour in both sets of
parametrization, the main difference is in the behaviour of $E^u$ and $E^d$ 
(see Fig. 9 of \cite{liuti}) at the input scale. $E^u$ is a slowly
increasing function of $x$ for set I, and for set II, it  increases rapidly
as $x$ approaches zero. $E_d$ has a peak at larger value of $x$ for set II. 
It was found that the difference between the sets decreases if one evolves
the GPDs to higher values of $Q^2$ (scale). 
It is to be noted that the parametrizations for \cite{marc} and \cite{diehl}
are at different input scales compared to \cite{liuti}. However the GPDs are
evolved to the respective scales of \cite{marc} and \cite{diehl} and a
comparison is provided in Figs. (13-16) of \cite{liuti}. For $H^u$ and $H^d$  
although there is agreement for low $-t$, for higher $-t$ the results differ
qualitatively and quantitatively. For $E^u$ and $E^d$ the disagreement 
is also at lower $-t$.

Again, even at $Q^2=4 ~\mathrm{GeV}^2$, there is significant difference between set I and set II.       
$H^u$ and $H^d$ agrees with lattice calculations of \cite{lattice} at
$t=-0.3 ~\mathrm{GeV}^2$, but the qualitative behaviour is different and even
is outside the error band of lattice calculations for higher values of $\mid
t \mid$. 

\vskip0.2in
\noindent
{\bf Parton distributions in impact parameter space}

Parton distribution in impact parameter space $q(x,b)$ is defined
as \cite{burkardt}:
\be
q(x,b)={1\over 4 \pi^2} \int d^2 \Delta e^{-i \Delta^\perp \cdot b^\perp}
H(x,t)\nonumber\\
e(x,b)={1\over 4 \pi^2} \int d^2 \Delta e^{-i \Delta^\perp \cdot b^\perp}
E(x,t).
\ee


These functions have the physical interpretation of measuring the
probability to find a quark of longitudinal momentum fraction $x$ at a
transverse position $b_\perp$ in  the nucleon.
Here $b= \mid b_\perp \mid $ is the impact parameter which is the 
transverse distance between
the struck parton and the center of momentum of the hadron.  $b$ is defined 
such a way that $\sum_i x_i b_i=0$ where the sum is over the number of partons.
The relative distance ${b\over 1-x}$ between the struck parton and the
spectator system provides an estimate of the size of the system as a whole.


In   Fig. 1-2 we have plotted $q(x,b)$ and $e(x,b)$,
both for u and d quarks and for set I.  The values of the
parameters used are from \cite{liuti} at the input scale. For small
and medium $x$, $e^d(x,b)$ is larger in magnitude than $e^u(x,b)$. The peak
shifts to higher $x$ as $b$ decreases. This means that the $d$ quark
dominates in the proton helicity flip distribution. However, $u$ quark
contribution dominates in the helicity non-flip $q(x,b)$. $e^d$ is negative
whereas $e^u$ is positive, similar to the model in \cite{marc}. However, in
the model we study, $e^d(x,b)$ is comparable with or even larger in magnitude 
than $q^d(x,b)$, unlike in \cite{marc}, where it is much smaller at the 
input scale. In the parametrization of \cite{marc}, the $t$
dependence is only in the argument of the exponential and the Fourier
transform is simpler and can be obtained analyically. The resulting parton
distributions in the impact parameter space are larger in magnitude than
what we get in this work using the parametrization of \cite{liuti} which may
be due to the different scale $Q_0^2$ in the plots. In
\cite{diehl}, $2-D$ distributions in the $b_x-b_y$ plane are plotted.  The
smearing of the quark distributions in the transverse impact parameter plane
decreases as $x$ increases, which means that the parton distributions are
more localized for higher values of $x$. Similar behaviour is observed in
the model of \cite{marc}. As $x$ approaches $1$, the
transverse width of $q(x,b)$ should vanish \cite{burkardt}. In this
limit $q(x,b)$ should have a very peaked transverse profile, as $H(x,t)$
is independent of $t$ when $x \rightarrow 1$ as the active quark carries all
the proton momentum no matter what $t$ is. However, from Fig. 1 we see that 
as $x \rightarrow 1$, the peak of the distribution decreases.

$q(x,b)$ both for u and d quarks in set II are the same as in set I.
In Figs. 3, we have plotted $e(x,b)$ for u and d quarks where
the parameters are as in set II. $e(x,b)$ has  a different behaviour
compared to set I. The peak of $e^u$ is shifted to 
very small value of $x$ and $e^u(x,b)$ decreases sharply as $x$ increases. 
That means at larger $x$, d quark dominates in $e(x,b)$.

The Fourier transform of the GPD $E(x,t)$ plays an important role when
instead of an unpolarized target we have a transversely polarized target. 
In other words, it has a probability interpretation in the transversity
basis rather than the helicity basis. For a state polarized in the x
direction, parton distribution in the impact parameter space becomes
\cite{burkardt}
\be
q^X(x,b)=q(x,b)-{1\over 2 M} {\partial e (x,b) \over \partial b_y} 
\ee
This means that the GPD $E(x,t)$ causes a transverse shift of the quark
distribution in a transversely polarized target. For a state polarized in x
direction the shift is in the y direction and so on. The magnitude of the 
shift is given by ${1\over 2 M} \mid {\partial e (x,b) \over \partial 
b_y} \mid $. The average displacement of the shift is given by
\be
{{\langle b^y \rangle}^q}_X={\int d^2b b^y q^X(x,b)\over \int d^2b q^X(x,b)} = {1\over 2 M}
{E^q(x,0)\over H^q(x,0)}.
\ee  


The distance between the struck quark and the spectator system is given by
\be
s^q(x)={{<b^y>}^q_X\over 1-x} 
\ee
For $d$ quarks, $s^q(x)$ is larger in magnitude than $u$ \cite{diehl}. The
transverse shift depends on the set of parameters used in the model
considered here \cite{liuti}. 

The parton distributions in impact parameter space should obey the
positivity condition \cite{burkardt2} 
\be
q(x,b) \ge {1\over 2 M} \mid {\partial e (x,b) \over \partial b_i} \mid.
\ee
This follows from the fact that $q^X(x,b)$ which is the unpolarized parton
distribution for transversely polarized proton should have probability
interpretation.

The positivity relation in effect puts an upper bound on  $b$. In the model
of \cite{marc}, it was found that for u quarks, the positivity bound was
satisfied over most of the $x$ region, considering that the GPDs are
vanishingly small for values of $b$ larger than the nucleon size. 
For d quarks, a violation of positivity was observed which is more
pronounced at larger values of $x$ and $b$. However the violation is small.
In the model of \cite{diehl}, the positivity condition was used to constrain
the behaviour of $E(x,t=0)$ at large values of $x$. In Fig. 4  we have
plotted $q(x,b)$ as well as ${1\over 2 M} \mid {\partial e (x,b)
 \over \partial b} \mid$ for u and d quarks for both the parametrizations in
\cite{liuti}, as functions of $b$. We see that in set I for u quarks,
positivity is mildly violated for very large values of $x$ and for $b > 0.5
~\mathrm{GeV}^{-1}$. For u quarks in set II there is no violation. 
For d quarks in set I, positivity is not violated for any $b$ values for
$x<0.1$. It is violated at larger $b$ values or comparatively larger $x$.
Violation is more as $x$ increases. For d quarks in set II, positivity is
largely violated for large $x,b$ values. 
To see the $x$ dependence of the violation of positivity, in Fig. 5 
we plot the same quantities as in Fig. 4, as  functions of $x$, for fixed
values of $b$. For u quarks in set I positivity is violated for $b \approx
0.4 ~\mathrm{GeV}^{-1}$ for large values of $x$. For higher $b$ values,
violation is even at smaller values of $x$. Violation of positivity is not
seen in u quarks in set II. For d quraks in set I, there is larger violation
of positivity and it starts already at $x =0.2$ for large $b$. For d quarks
in set II there is again a large violation of positivity that starts at $
x \approx 0.3$ for $b =5 ~\mathrm{GeV}^{-1}$.  
    
\vskip0.2in
\noindent
{\bf Conclusion}

In this paper, we have studied the GPDs $H(x,t)$ and $E(x,t)$ for zero
skewness in  a recently parametrized form \cite{liuti} in transverse
position or impact parameter space. We present a comparative study between
several models. A violation of the positivity condition was observed in
certain range of $x,b$. This puts additional constraint on the kinematical
region where this parametrization is to be used. This depends on the set of
parameters used in the model. A new fit of the parameters with this
additional constraint may improve the model. As extension to nonzero $\zeta$
was proposed in  \cite{l2}. In  a future work, we plan to investigate the
GPDs with nonzero skewness both in  transverse and longitudinal position
spaces. In particular, a study in longitudinal position space is interesting
to understand  the origin of the observed diffraction pattern  in the DVCS amplitude in a
simple QED model \cite{hadron_optics}. 

\vskip0.2in
\noindent
{\bf Ackowledgement:} 
AM thanks DST fasttrack scheme, Govt. of India, for support. We thank S.
Liuti and S. Ahmad for helpful communication.


 \newpage

\begin{figure}[!htp]
\begin{minipage}[c]{0.9\textwidth}

\tiny{(a)}\includegraphics[width=6.5cm,height=5cm,clip]{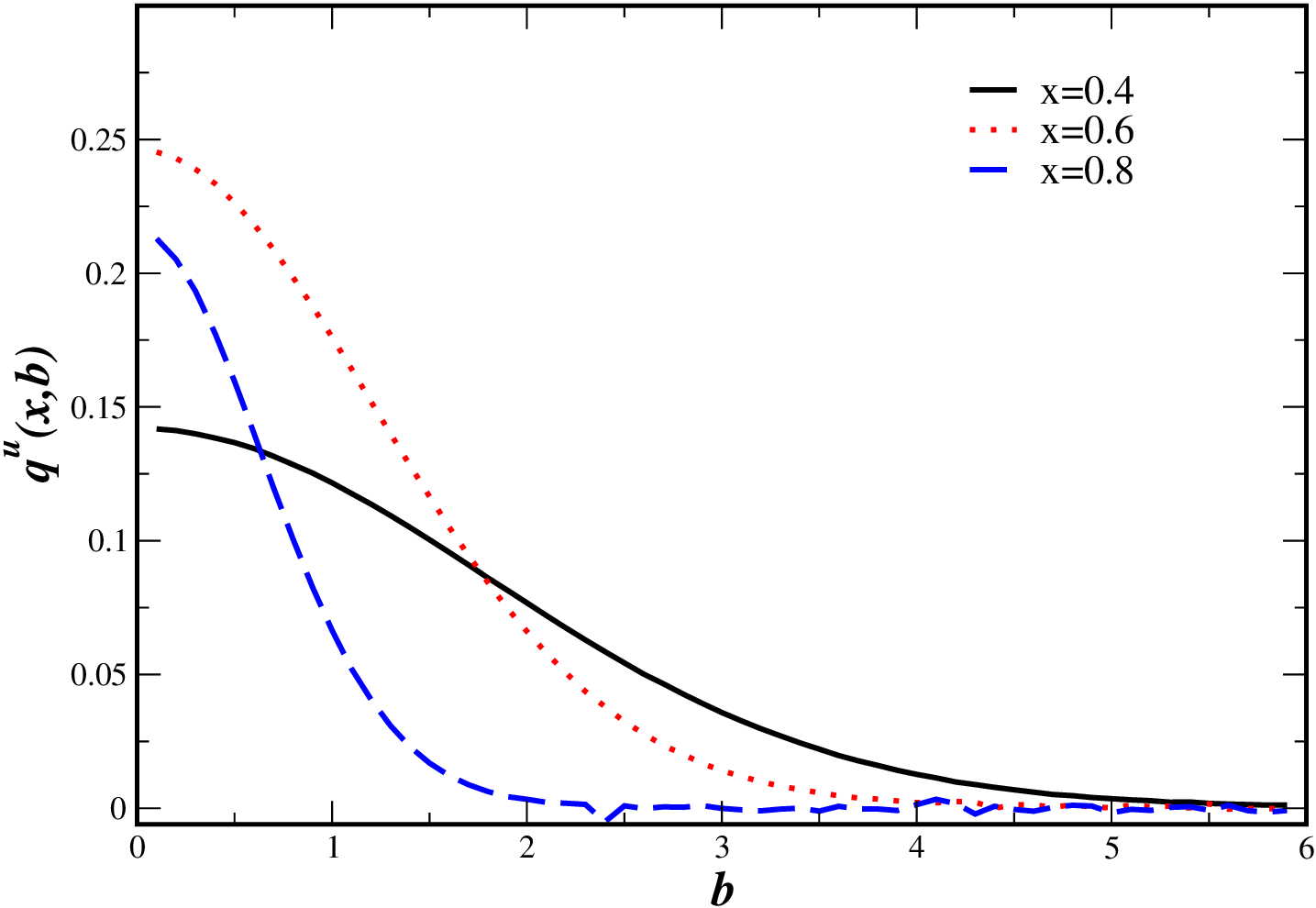}
\hspace{0.1cm}
\tiny{(b)}\includegraphics[width=6.5cm,height=5cm,clip]{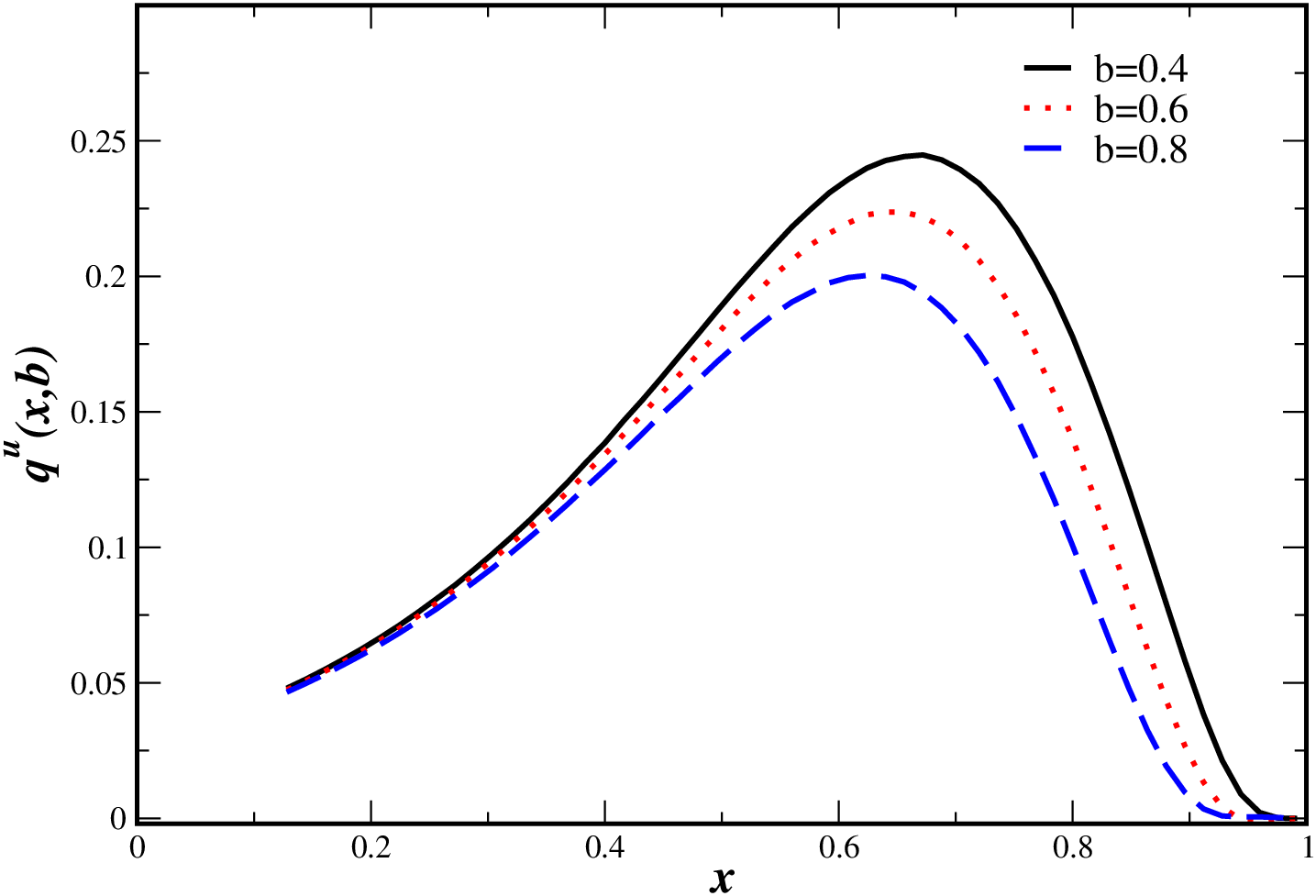}
\end{minipage}
\begin{minipage}[c]{0.9\textwidth} 
\tiny{(c)}\includegraphics[width=6.5cm,height=5cm,clip]{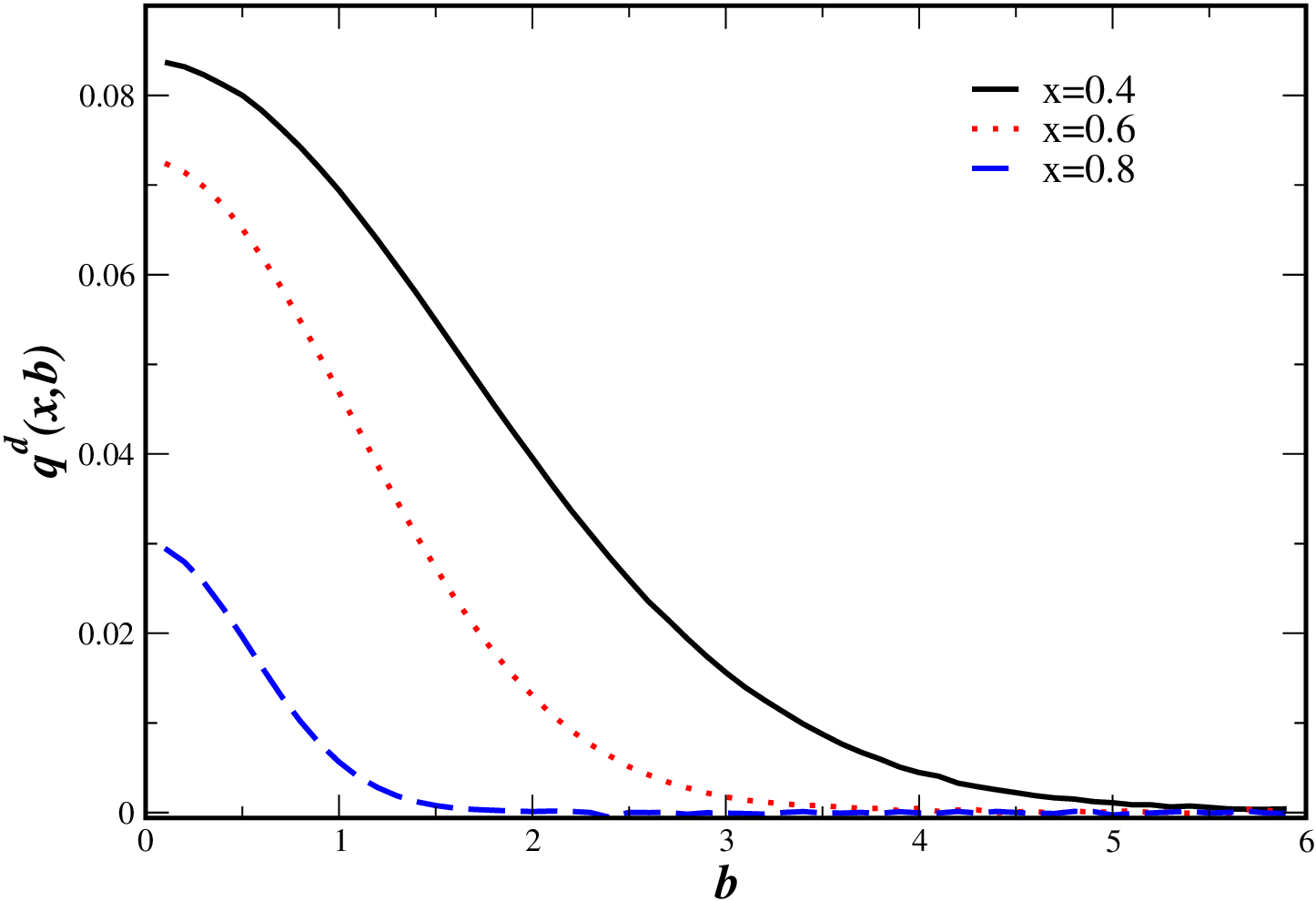}
\hspace{0.1cm}%
\tiny{(d)}\includegraphics[width=6.5cm,height=5cm,clip]{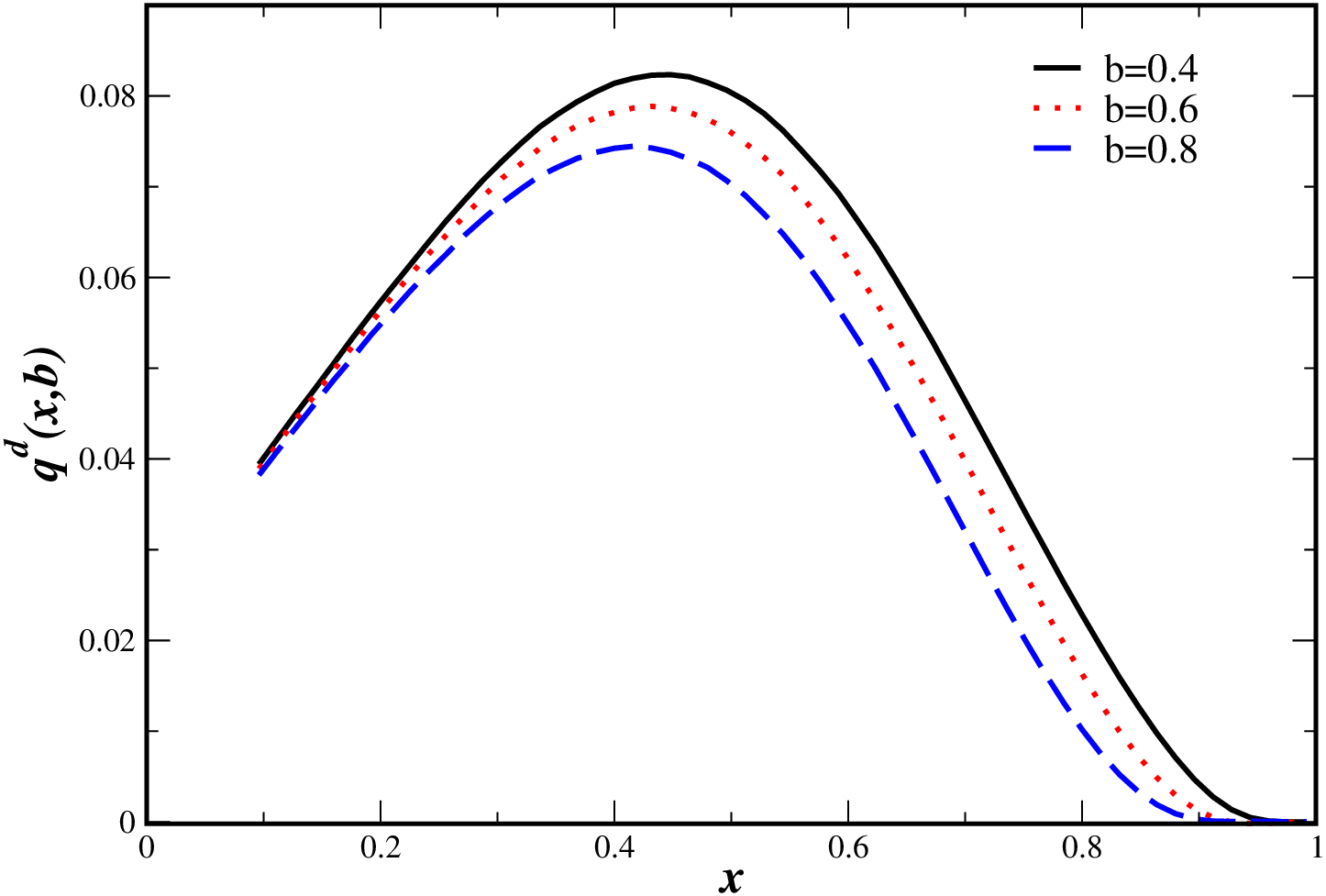}
\end{minipage} 
\caption{\label{s1_h} (Color online) Plots of (a) $q^u (x,b)$ vs $b= \mid
b_\perp \mid$ for fixed values of $x$, (b) $q^u (x,b)$ vs $x$ for 
fixed values of $b$, (c) same as in (a) but for $q^d$, (d) 
same as in (b) but for $q^d$. Parameters are as in set I. $b$ is in
$\mathrm{GeV}^{-1}$.}
\end{figure}

\begin{figure}[!htp]
\begin{minipage}[c]{0.9\textwidth} 
\tiny{(a)}{\includegraphics[width=6.5cm,height=5cm,clip]{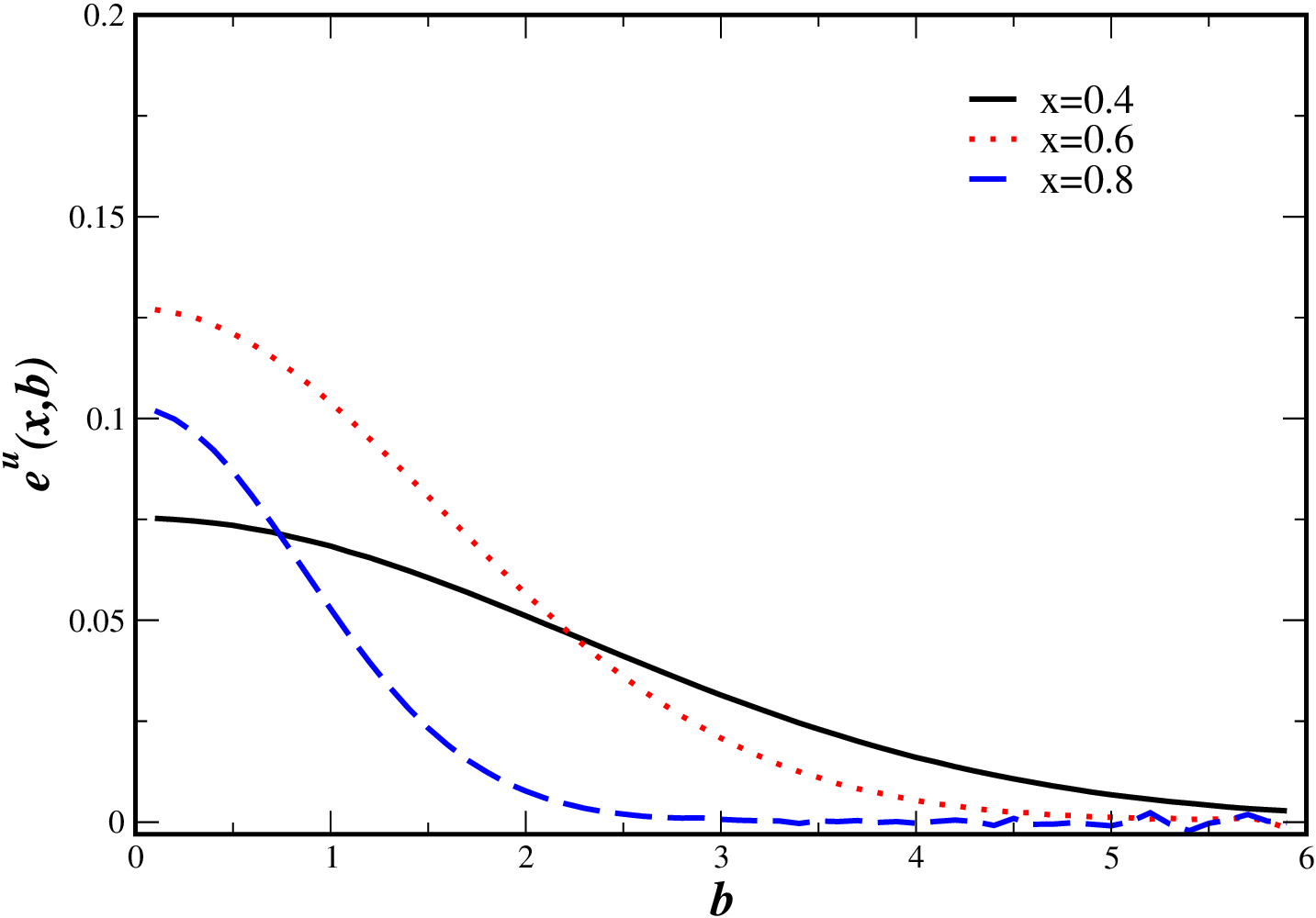}}
\hspace{0.1cm}
\tiny{(b)}{\includegraphics[width=6.5cm,height=5cm,clip]{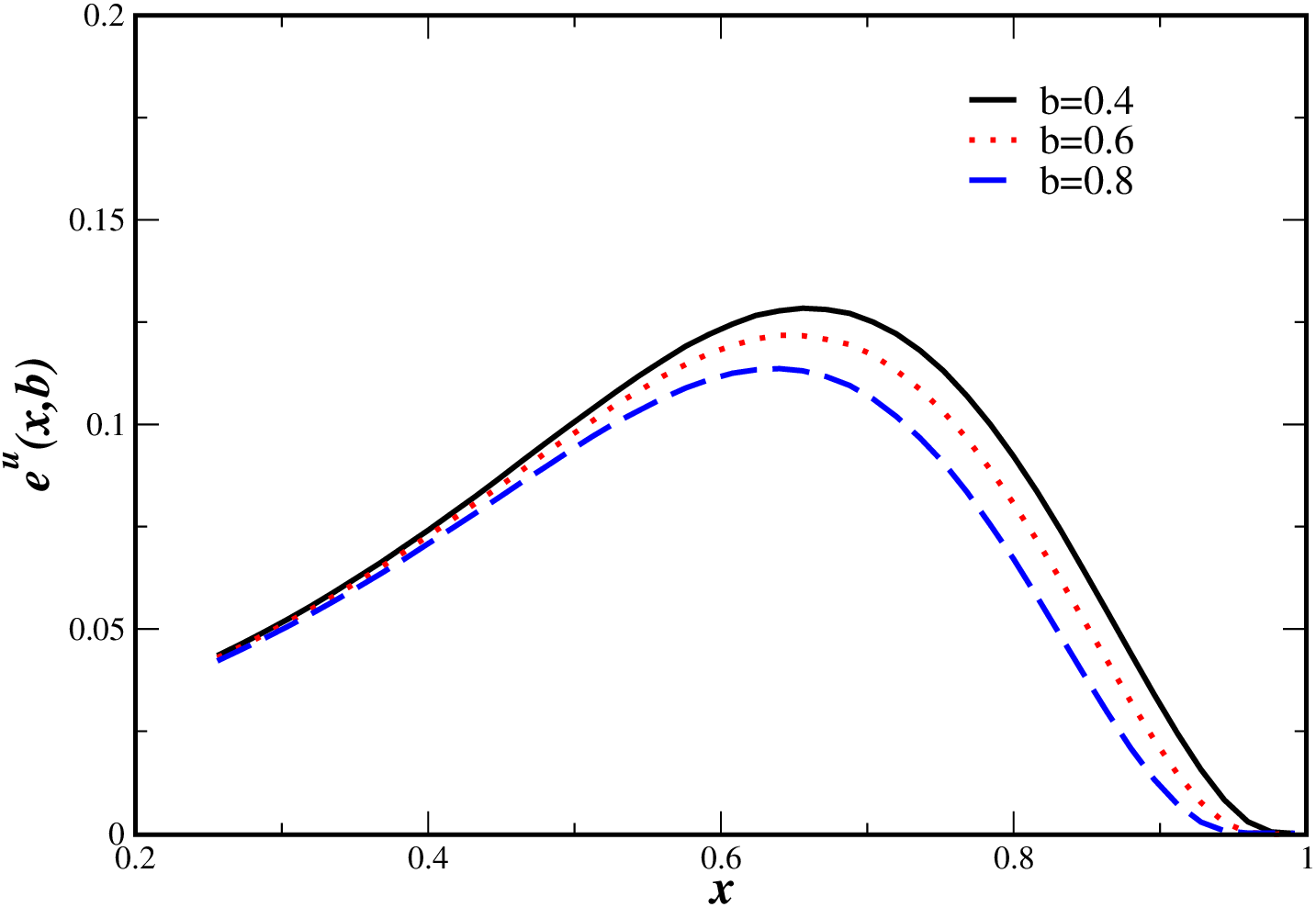}}
\end{minipage}
\begin{minipage}[c]{0.9\textwidth} 
\tiny{(c)}\includegraphics[width=6.5cm,height=5cm,clip]{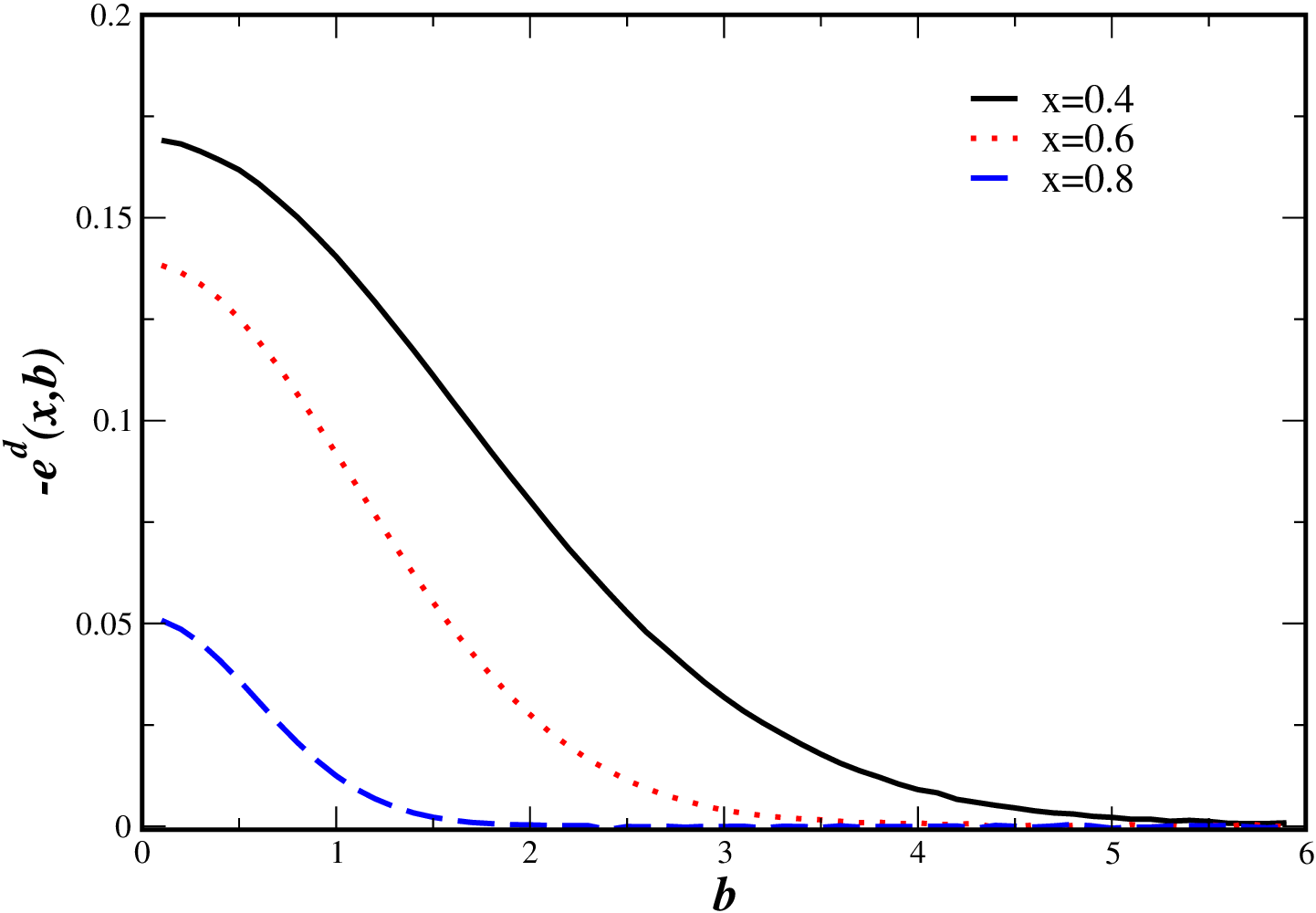}
\hspace{0.1cm}%
\tiny{(d)}\includegraphics[width=6.5cm,height=5cm,clip]{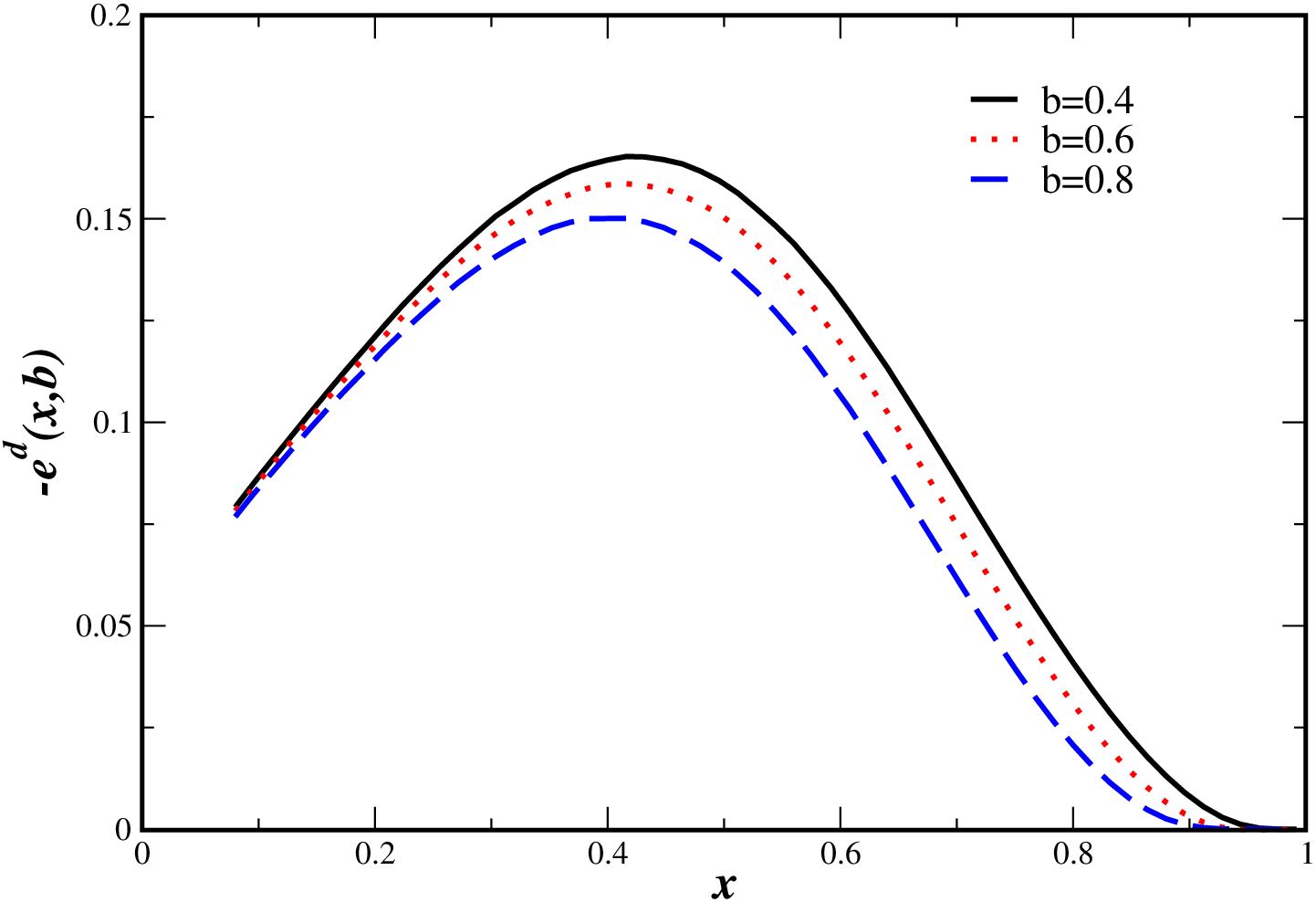}
\end{minipage} 
\caption{\label{s1_e} (Color online) Plots of (a) $e^u (x,b)$ vs $b= \mid
b_\perp \mid$ for fixed values of $x$, (b) $e^u (x,b)$ vs $x$ for 
fixed values of $b$, (c) same as in (a) but for $-e^d$, (d) 
same as in (b) but for $-e^d$. Parameters are as in set I. $b$ is in
$\mathrm{GeV}^{-1}$.}
\end{figure}
\begin{figure}[!htp]
\begin{minipage}[c]{0.9\textwidth} 
\tiny{(a)}\includegraphics[width=6.5cm,height=5cm,clip]{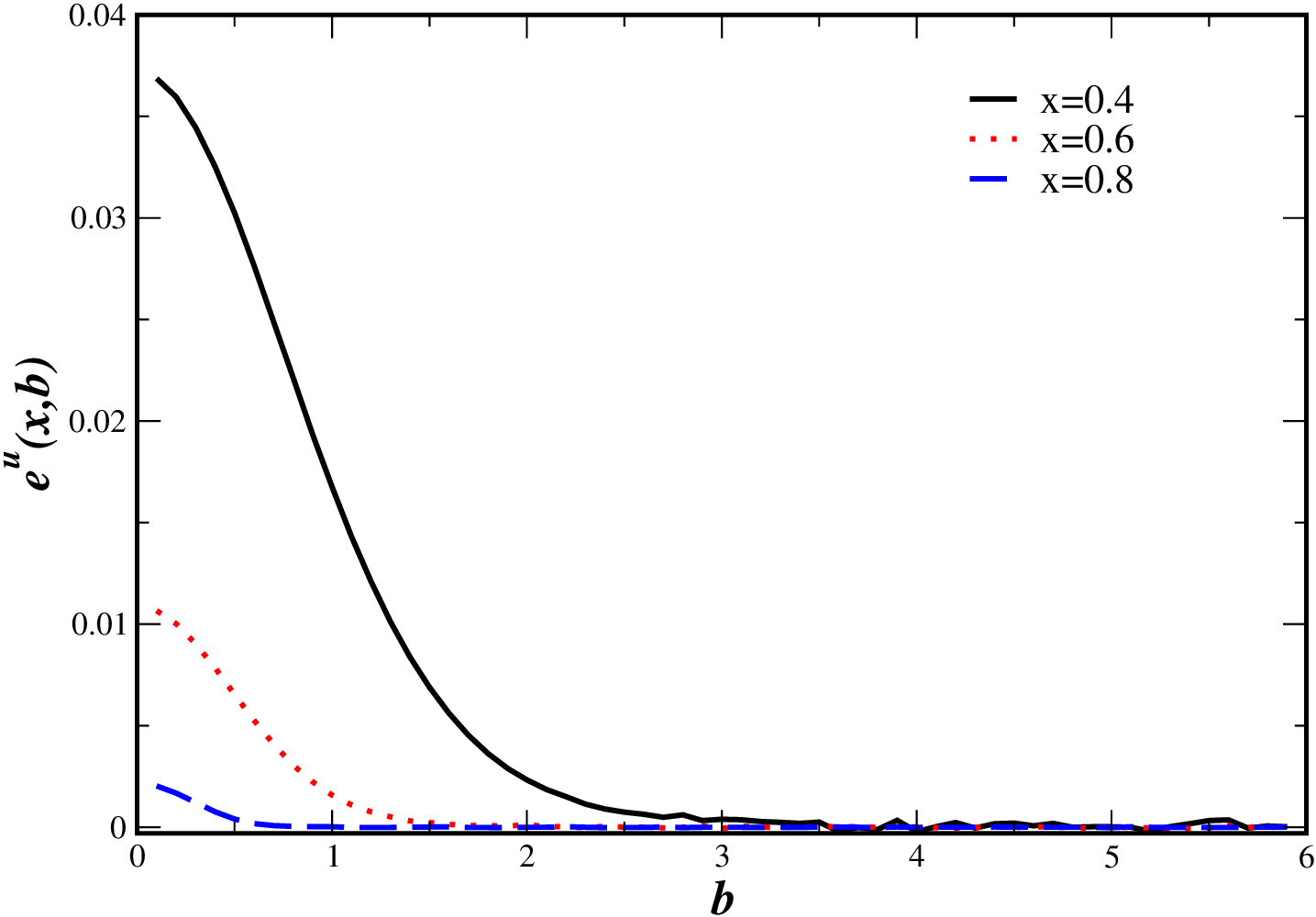}
\hspace{0.1cm}%
\tiny{(b)}\includegraphics[width=6.5cm,height=5cm,clip]{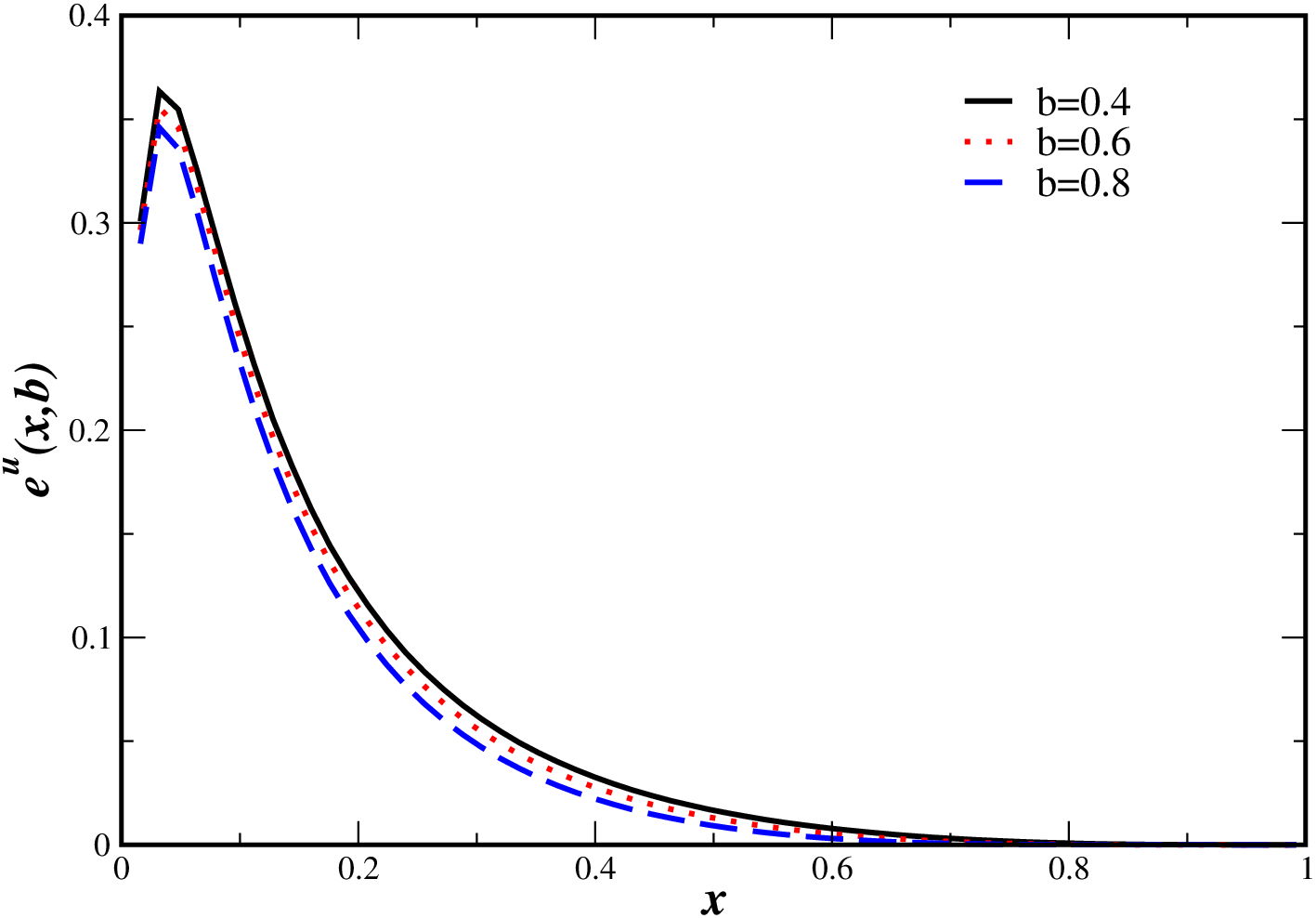}
\end{minipage}
\begin{minipage}[c]{0.9\textwidth} 
\tiny{(c)}\includegraphics[width=6.5cm,height=5cm,clip]{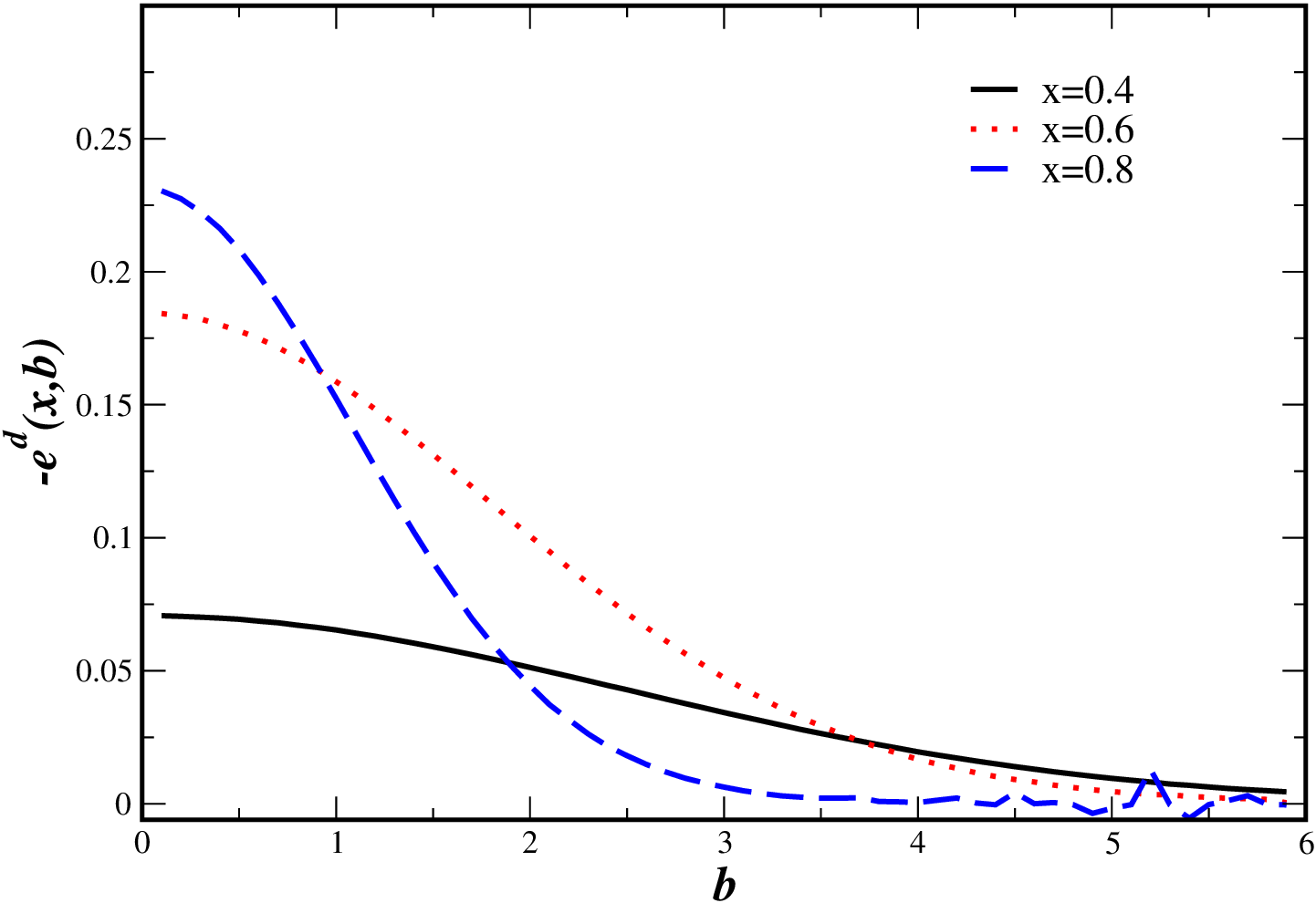}
\hspace{0.1cm}%
\tiny{(d)}\includegraphics[width=6.5cm,height=5cm,clip]{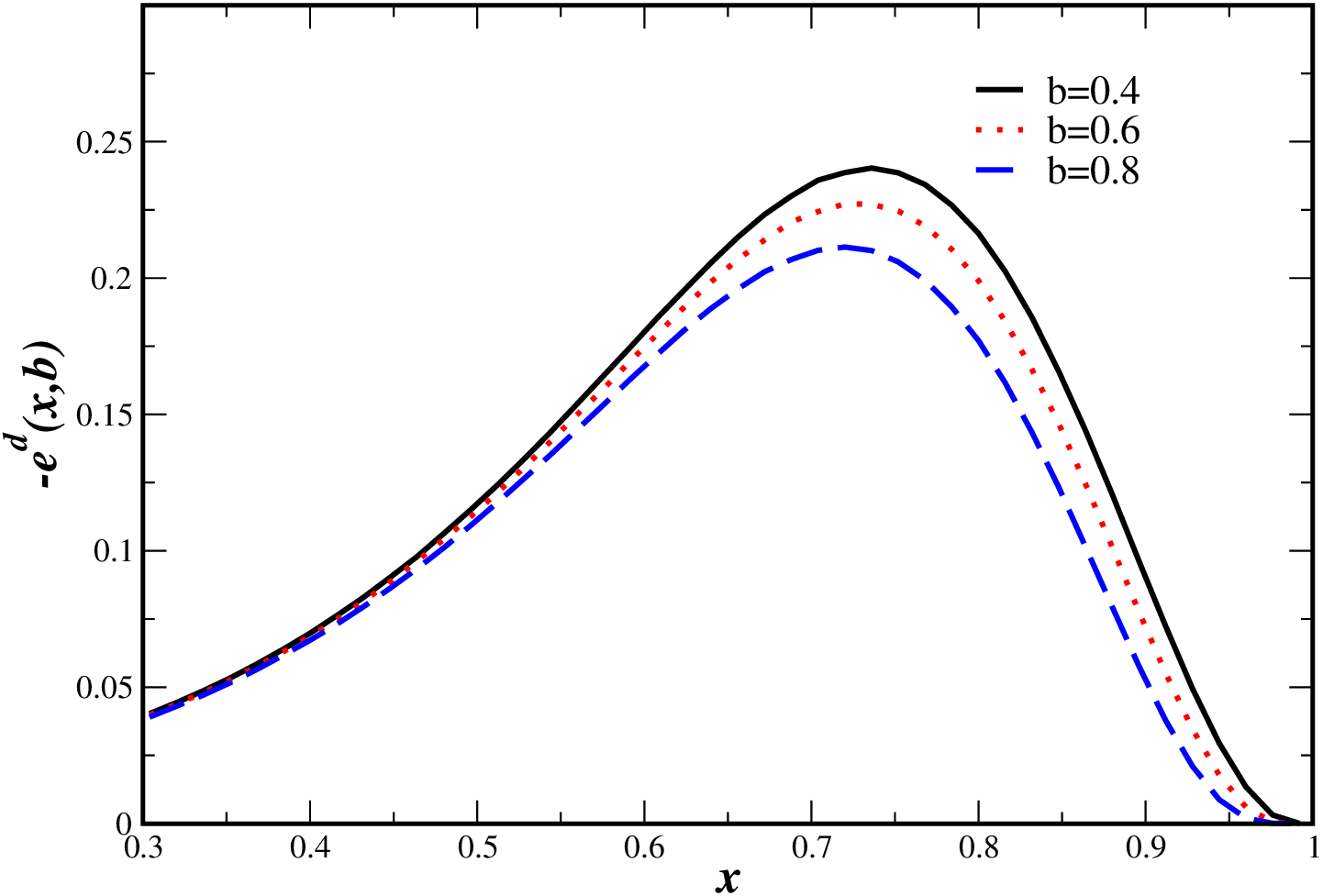}
\end{minipage}
\caption{\label{s2_huvsbx}(Color online)  Plots of (a) $e^u (x,b)$ vs $b= \mid
b_\perp \mid$ for fixed values of $x$, (b) $e^u (x,b)$ vs $x$ for 
fixed values of $b$, (c) same as in (a) but for $-e^d$, (d) 
same as in (b) but for $-e^d$. Parameters are as in set II. $b$ is in
$\mathrm{GeV}^{-1}$. }
\end{figure}
\begin{figure}[!htp]
\begin{minipage}[c]{0.9\textwidth} 
\tiny{(a)}\includegraphics[width=6.5cm,height=5cm,clip]{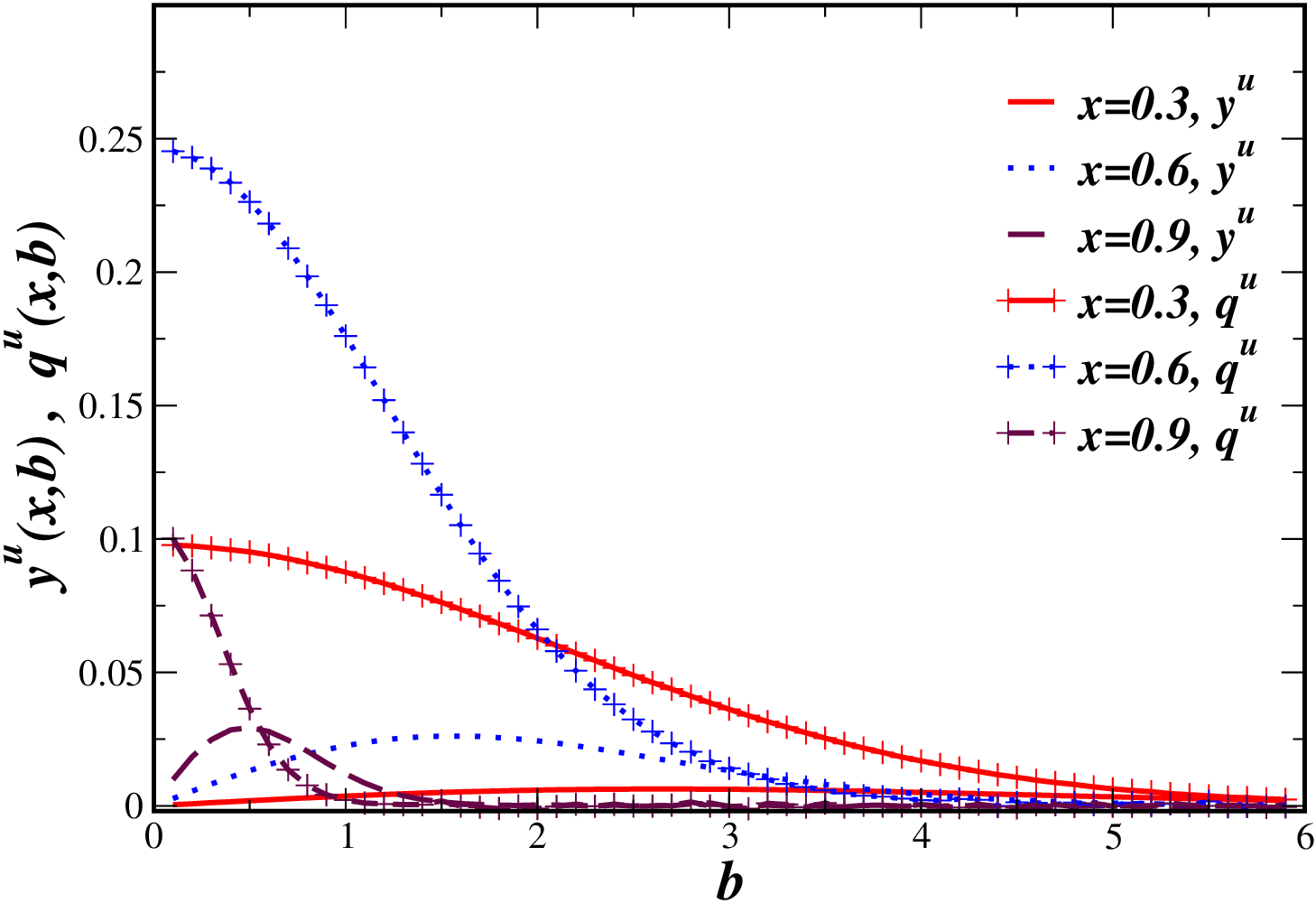}
\hspace{0.1cm}%
\tiny{(b)}\includegraphics[width=6.5cm,height=5cm,clip]{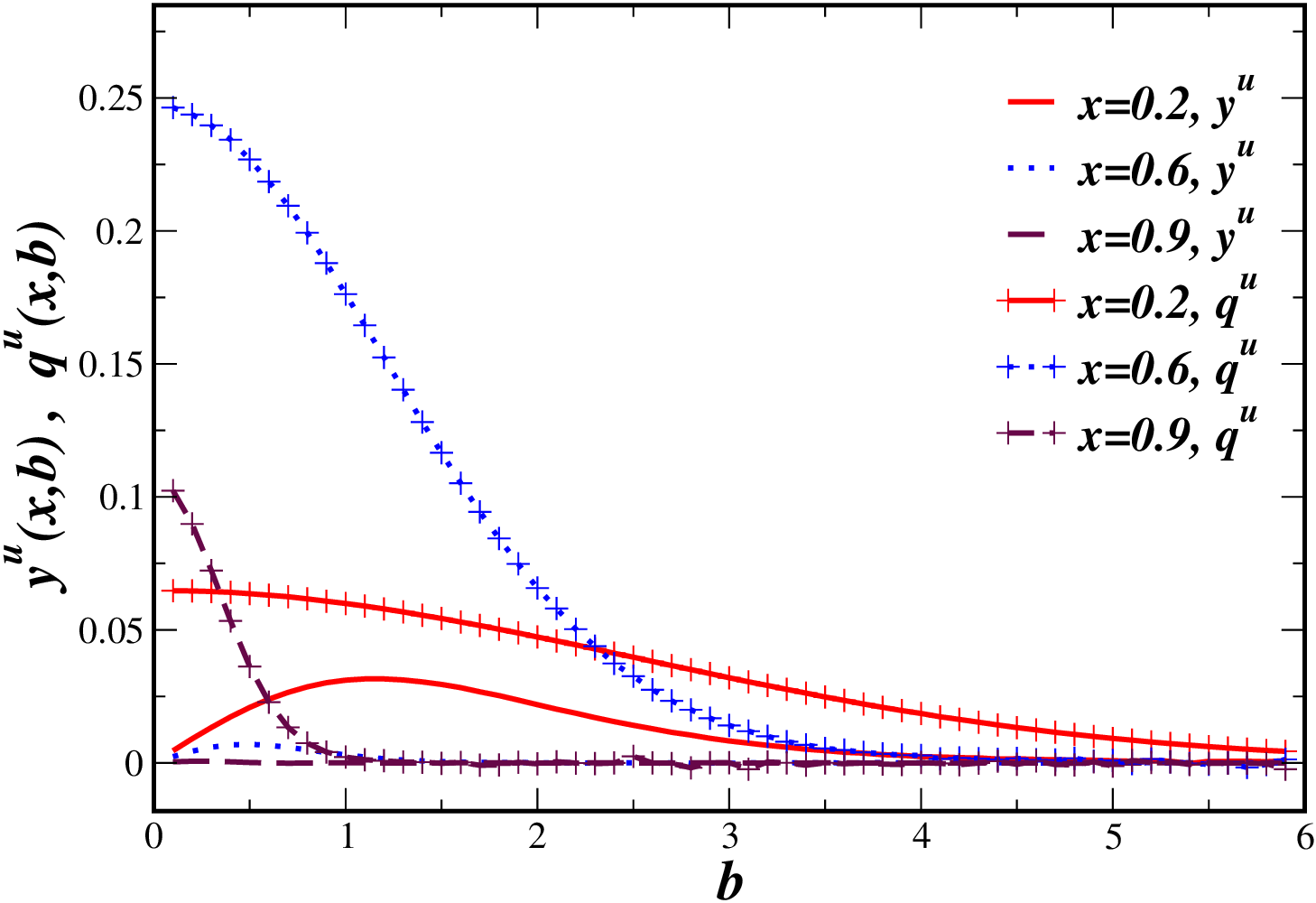}
\end{minipage} 
\begin{minipage}[c]{0.9\textwidth} 
\tiny{(c)}\includegraphics[width=6.5cm,height=5cm,clip]{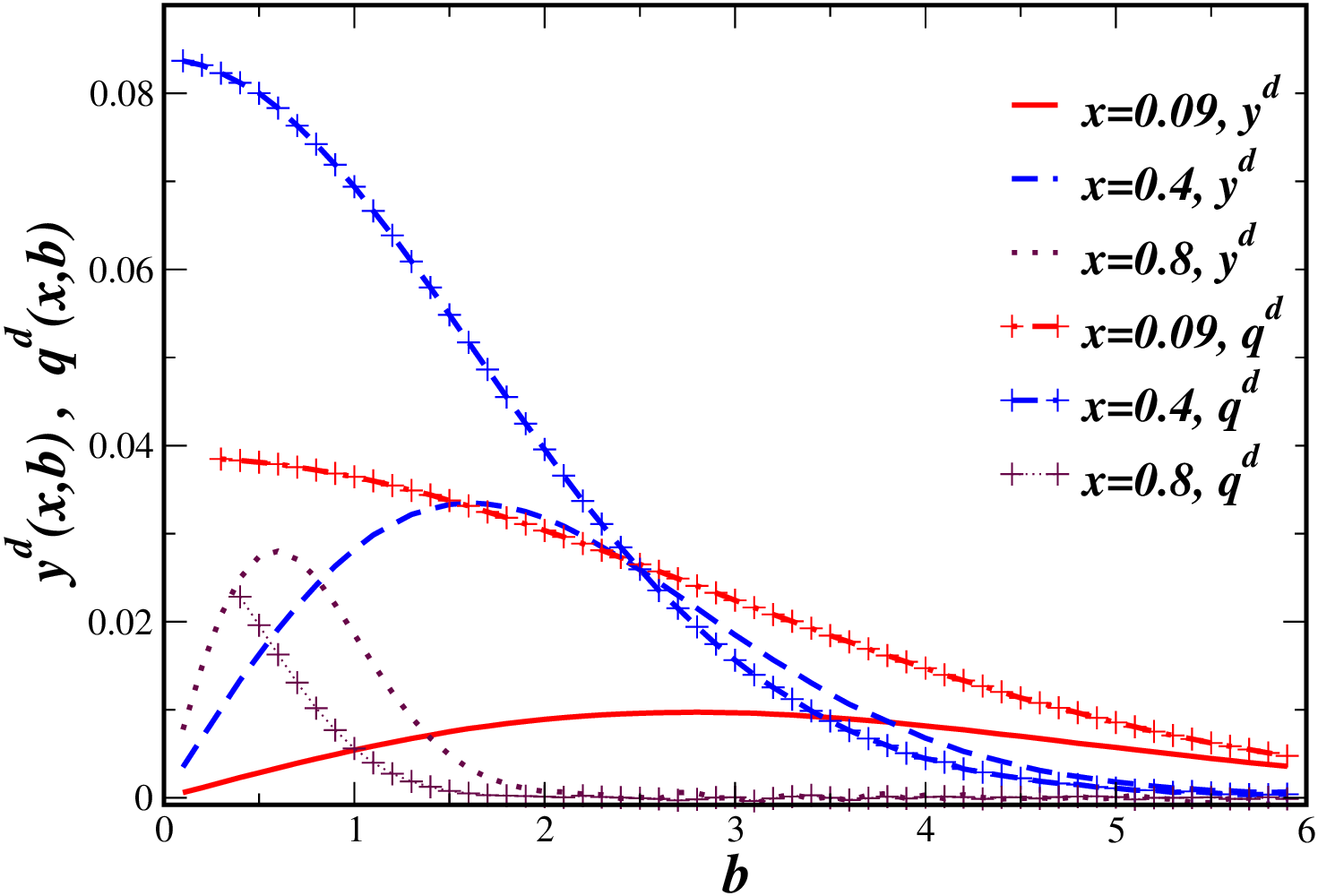}
\hspace{0.1cm}%
\tiny{(d)}\includegraphics[width=6.5cm,height=5cm,clip]{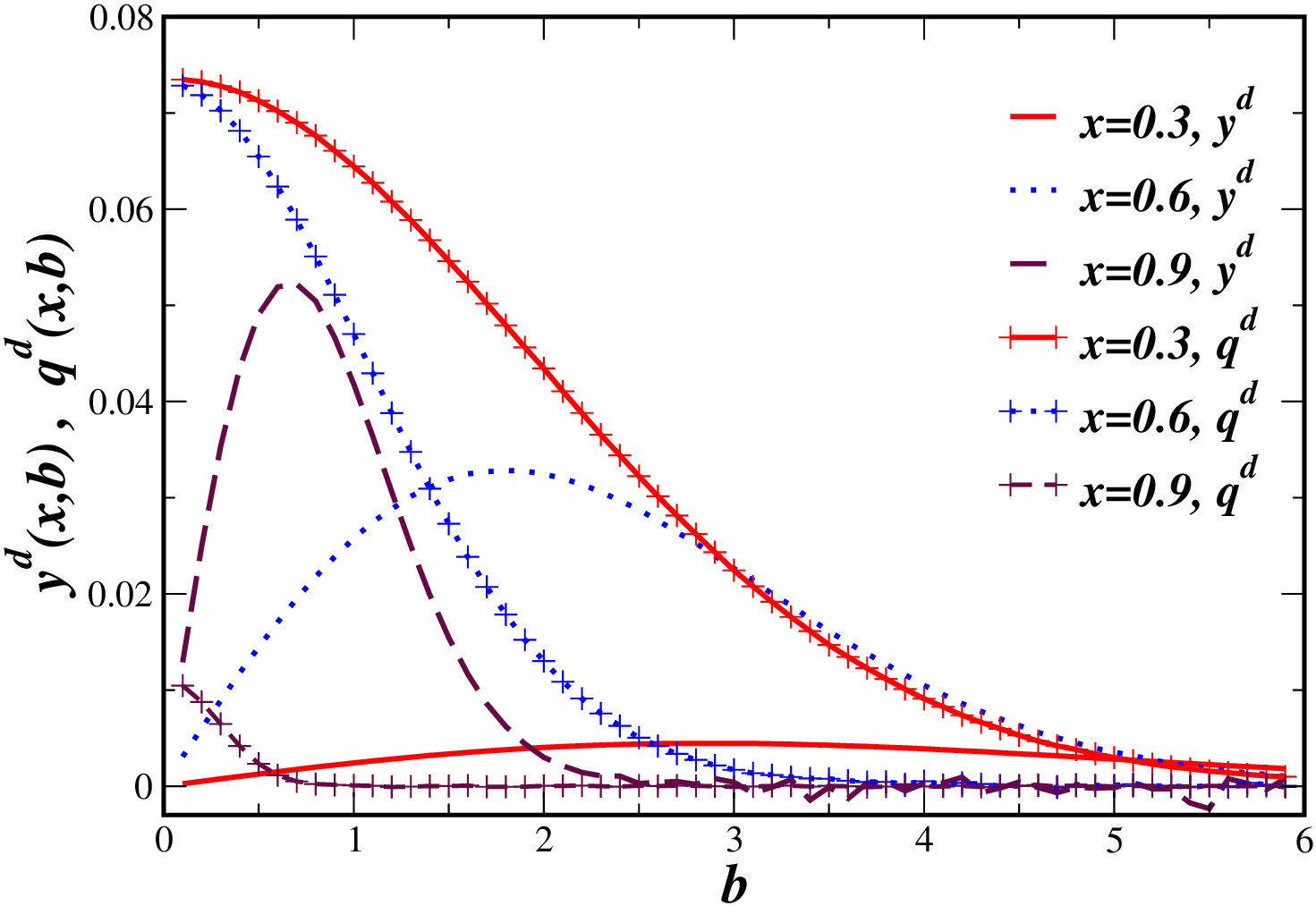}
\end{minipage} 
\caption{\label{ineqb}(Color online)  Plots of  (a) $q^u(x,b)$ and $y^u(x,b)=
{1\over 2 M} \mid {\partial e^u(x,b) \over \partial b} \mid$ vs. $b$ for set
I, (b) the same as in (a) but for set II, (c) $q^d(x,b)$ and $y^d(x,b)=
{1\over 2 M} \mid {\partial e^d(x,b) \over \partial b} \mid$ vs. $b$ for set
I, (d) the same as in (c) but for set II. $b$ is in
$\mathrm{GeV}^{-1}$.}      
\end{figure}
\begin{figure}[!htp]
\begin{minipage}[c]{0.9\textwidth} 
\tiny{(a)}\includegraphics[width=6.5cm,height=5cm,clip]{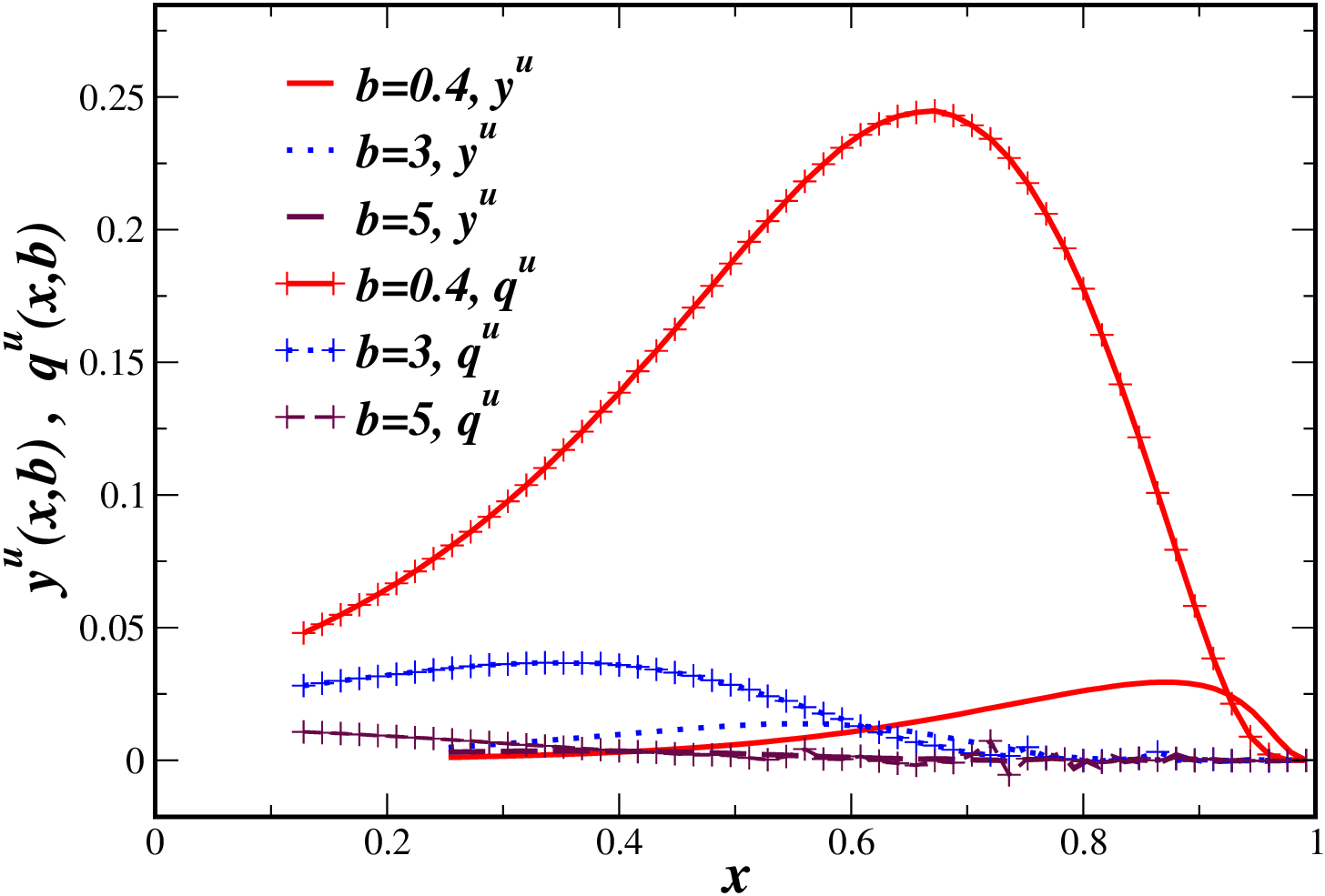}
\hspace{0.1cm}%
\tiny{(b)}\includegraphics[width=6.5cm,height=5cm,clip]{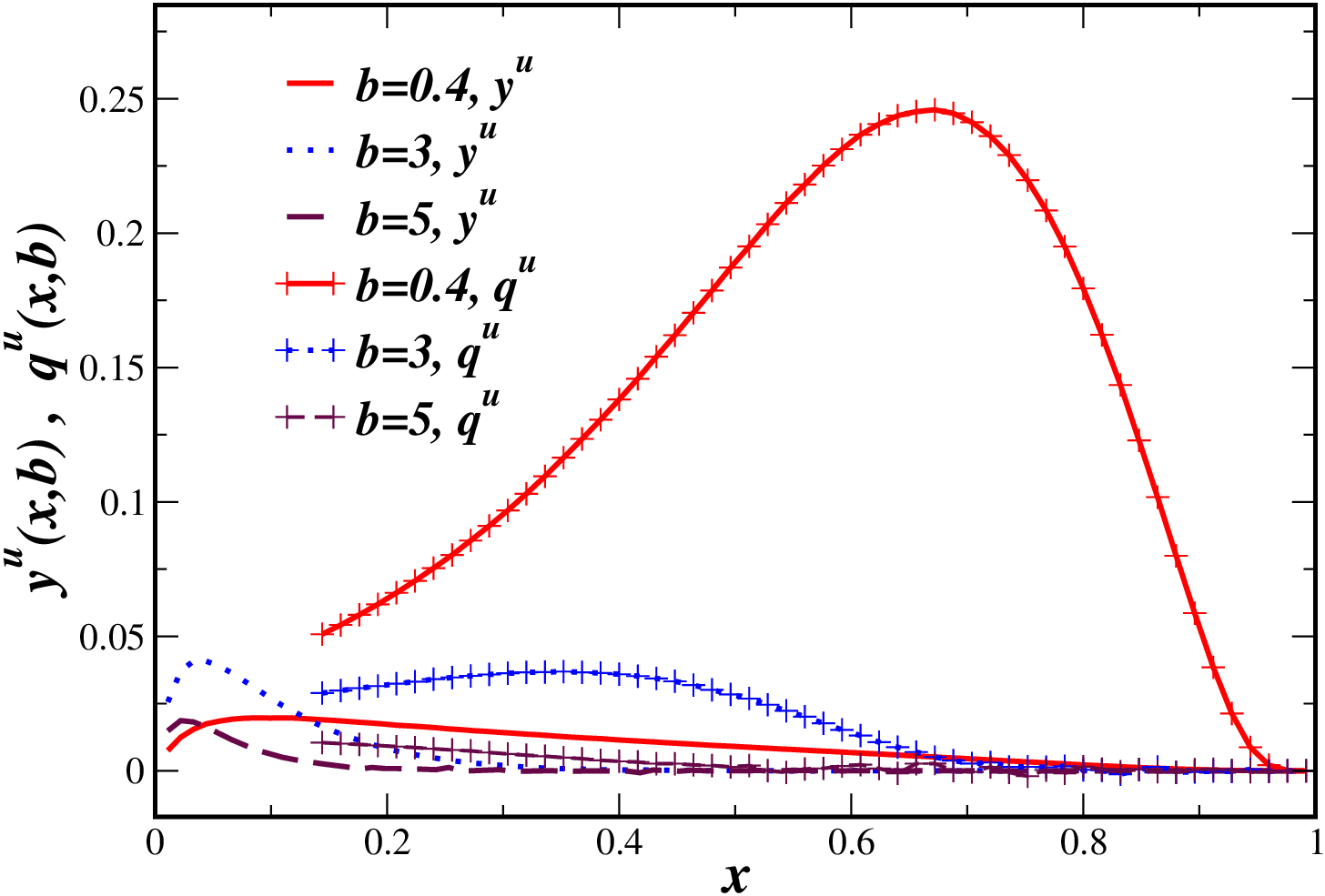}
\end{minipage} 
\begin{minipage}[c]{0.9\textwidth} 
\tiny{(c)}\includegraphics[width=6.5cm,height=5cm,clip]{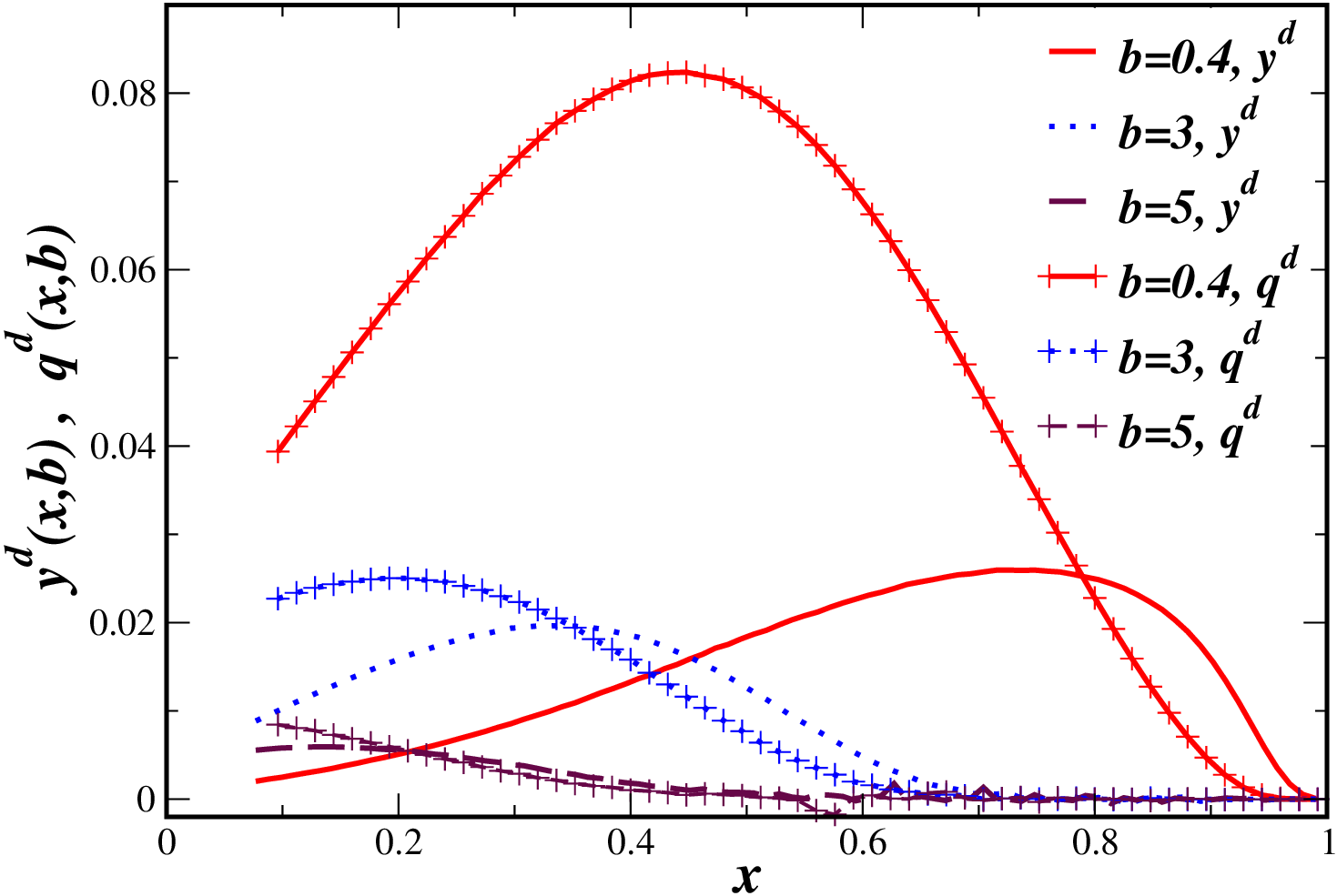}
\hspace{0.1cm}%
\tiny{(d)}\includegraphics[width=6.5cm,height=5cm,clip]{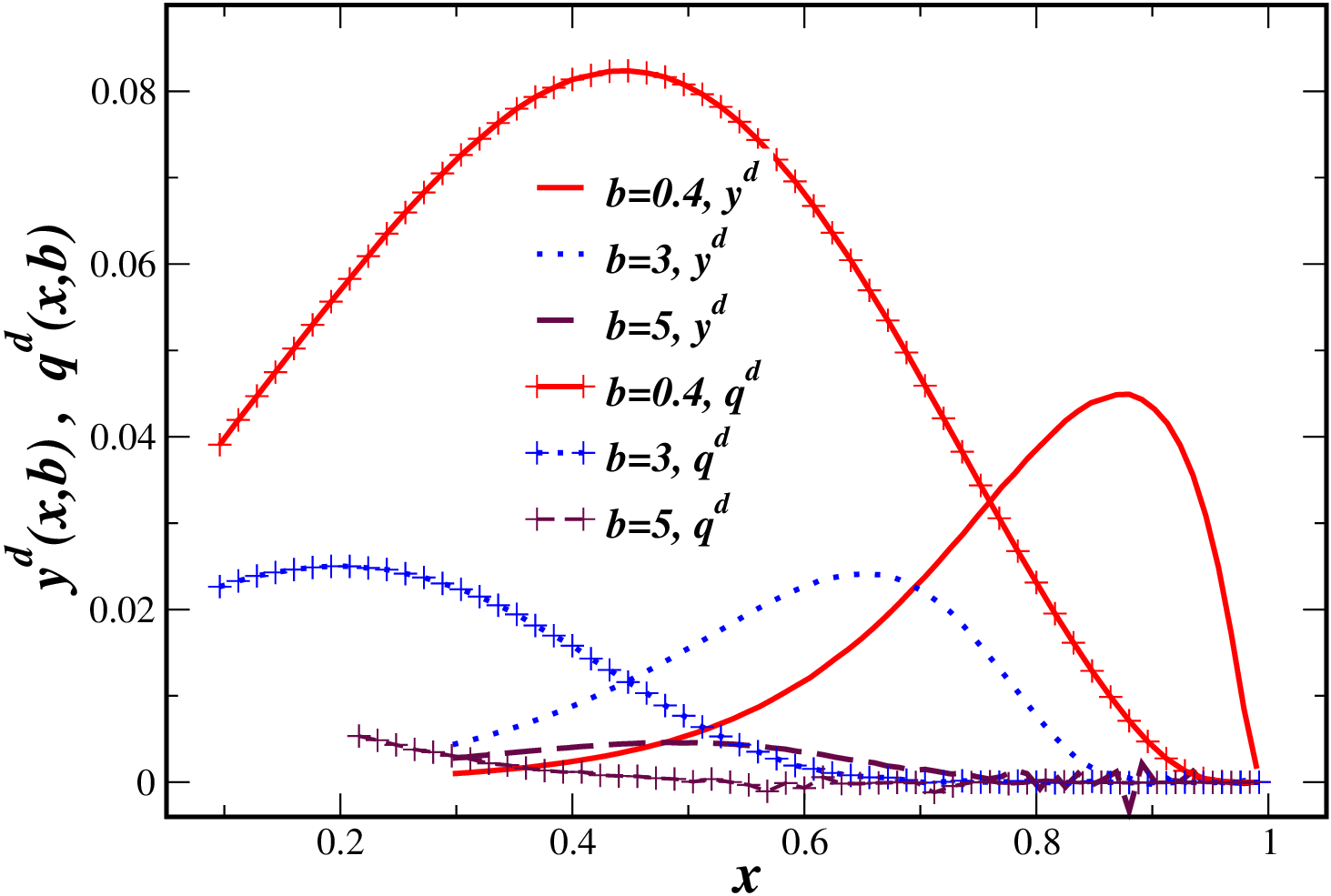}
\end{minipage} 
\caption{\label{ineqx}(Color online)  Plots of  (a) $q^u(x,b)$ and $y^u(x,b)=
{1\over 2 M} \mid {\partial e^u(x,b) \over \partial b} \mid$ vs. $x$ for set
I, (b) the same as in (a) but for set II, (c) $q^d(x,b)$ and $y^d(x,b)=
{1\over 2 M} \mid {\partial e^d(x,b) \over \partial b} \mid$ vs. $x$ for set
I, (d) the same as in (c) but for set II. $b$ is in
$\mathrm{GeV}^{-1}$.}      
\end{figure}
\end{document}